\documentclass[11pt]{iopart}
 \expandafter\let\csname equation*\endcsname\relax
 \expandafter\let\csname endequation*\endcsname\relax

\usepackage{amssymb, amsmath}
\usepackage{graphicx,color}

\begin{document}

\title[Distribution of zeros of the S-matrix of chaotic cavities]{Distribution of zeros of the S-matrix of chaotic cavities with localized losses and Coherent Perfect Absorption: non-perturbative results}
\vskip 0.2cm

\author{Yan V. Fyodorov}
 \address{King's College London, Department of Mathematics, London  WC2R 2LS, United Kingdom}
 \author{Suwun Suwunnarat and Tsampikos Kottos}
 \address{Department of Physics, Wesleyan University, Middletown, Connecticut 06459, USA}

\date{27th March 2017}

\begin{abstract}
We employ the Random Matrix Theory framework to calculate the density of zeroes of an $M$-channel scattering matrix describing a chaotic
cavity with a single localized absorber embedded in it. Our approach extends beyond the weak-coupling limit of the cavity with the channels and
applies for any absorption strength. Importantly it provides an insight for the optimal amount of loss needed to realize a chaotic coherent perfect
absorbing (CPA) trap. Our predictions are tested against simulations for two types of traps: a complex network of resonators
and quantum graphs.
\end{abstract}

\maketitle
One of the undisputed successes of the Random Matrix Theory (RMT) in last decades is its faithful description of scattering properties
of disordered or/and chaotic cavities\footnote{In the standard terminology of the field of quantum chaos a scattering domain is called chaotic (respectively, integrable) if after detaching all external leads or waveguides the dynamics of a classical particle bouncing inside the associated closed domain is chaotic (resp.  integrable). The generally accepted conjecture is that highly excited eigenstates of the Laplacian operator on the domains corresponding to chaotic cavities are faithfully described by eigenvalues of large random matrices, whereas eigenspectra of integrable cavities follow very different (Poisson) statistics.}  verified in a number of experiments: distribution of conductances, delay times,
resonance widths, 
 the consequences of preserved/violated time reversal symmetry, the presence of
uniform absorption etc, see e.g. recent review articles ~\cite{RMTscatt,kuhl13,grad14,diet15,hcao15}.

Inspired by all these successes 
 it is natural to expect that RMT should play a role among essential tools employed for the understanding
of a new category of scattering problems, associated with the design of chaotic coherent perfect absorbers (CPA).
CPAs are typically weakly lossy cavities which act as perfect constructive interference
traps for incident coherent radiation \cite{CGCS10}. CPA protocols \cite{CGCS10,L10} have been found increasingly relevant because of
our capabilities in manipulating incoming wavefront shapes in frameworks ranging from acoustics, to microwaves and optics.
Technological applications of CPAs might cover areas as diverse as optical light-light control and switching, plasmonic sensors, lasing/anti-lasing cavities \cite{CGCS10,WCGNSC11,ZFYZSF16,apl1,apl2,apl3} to
THz and radiofrequency wave trapping and stabilization of wireless communications \cite{apl6,SLLREK12} and sound absorption \cite{GTRMTP16}.   Despite
the broad range of technological applications the study of CPA was confined up to now to the investigation of  simple (integrable) cavities.
Only recently researchers started investigating the effects of complexity (chaos) in the realization of CPAs \cite{Tsampikos}. This is quite
surprising since at the heart of CPA protocols are wave interference phenomena which are abundantly  present in wave chaotic cavities. In
fact one can further capitalize on these complex (chaotic) interferences in order to provide {\it universal} predictions for properties
of chaotic CPA cavities.

In this paper we are utilizing the RMT toolbox in order to shed new light on the realization of chaotic CPAs with $M$ open channels and
spatially non-uniform losses.
 For simplicity we shall assume that the losses are described by a single point-like absorber  with loss-strength
$\gamma_0$ embedded inside the cavity. Specifically we provide a detailed statistical description of the density of complex zeros  of the scattering matrix $S$  ( i.e. solutions of $\det S = 0$ in the complex energy plane) which describes such chaotic CPA cavities, for a given strength of absorber $\gamma_0$. In particular,  the number of those  $S$-matrix zeroes  which with increased $\gamma_0$ move away from the upper-half plane and cross the real energy axis can be interpreted as the number of ``perfect absorbing states'' supported by such cavities. We provide analytical expressions
for the density of zeros as a function of $\gamma_0$ for a fixed $M$ and given universality class (violated/preserved time-reversal
invariance). We point out that CPAs associated with
violated time-reversal symmetry are novel and do not fall in the original category of CPAs
as defined in \cite{CGCS10,L10,WCGNSC11} the latter are typically recognized as the time-reversal of an outgoing
lasing field. Nevertheless, this new family of incident coherent fields lead to perfect absorption
as well.

In contrast to \cite{Tsampikos} we do not assume a weak-coupling limit,  thus providing {\it non-perturbative} insights
into the optimal amount of loss (perfect tuning) needed to realize a crossover of a zero from positive to negative complex energy
semi-plane. Our results are tested against detailed numerical simulations using both RMT modeling and an actual wave chaotic network (quantum graphs).

The main object of interest is a (non-unitary) $M$-channel scattering matrix $S\left(E,\gamma_0\right)$ which,  in the presence of  localized absorbers of strength $\gamma_0$, connects incoming $|I\rangle$ to outgoing $|O\rangle$ wave amplitudes at a fixed real energy $E$ via the relation $|O\rangle=S|I\rangle$. We start by a brief demonstration of the adjustments needed to be done in the approach of \cite{verb85,soko89,fyod97} in order to make it capable of describing the CPA problems.

In fact, one may conveniently think of a scattering system with $M\ge 1$ open scattering channels and $L\ge 1$ local absorbers of equal strength ( generalization to different strengths is obvious) as represented by a fictitious system of ${\cal M}=M+L$ open channels, such that its corresponding ${\cal M}\times {\cal M}$ scattering matrix ${\cal S}$  is unitary due to the flux conservation.  The building blocks for the associated ${\cal S}$-matrix are
(i) an $N\times N$ ($N\gg 1$) random Gaussian matrix $H$ (real symmetric GOE, $\beta=1$ or Hermitian GUE, $\beta=2$) used to model the
Hamiltonian (or, in context of classical waves, the wave operator) of an isolated chaotic cavity, (ii) an $N\times M$ matrix $W$ (containing the
{\it coupling amplitudes} to $M$ open scattering channels) satisfying, for simplicity, the relation $\sum_{l=1}^N W_{cl}^{*}W_{bl}=\gamma \delta_{cb}, \, \forall c,b=1,\ldots, M $ ('equivalent orthogonal channels assumption') and finally (iii) $N\times L$ matrix ${\cal A}$ of  coupling amplitudes $A_{li}$ describing  $L$ {\it absorptive} channels and taken as another set of fixed orthogonal vectors satisfying $\sum_{i=1}^N A_{li}A_{mi}=\gamma_0\delta_{cd}, \, \,\,\forall (l,m)=1,\ldots, L$.
We also assume that the vector space spanned by coupling vectors ${\cal A}_l, \,\, l=1,\ldots, L$ for the absorptive part is
orthogonal to one spanned by vectors $W_{b}, b=1,\dots M$. In such an approach the (sub-unitary) scattering matrix $S(E,\gamma_0)$ defined above is simply a $M\times M$ diagonal sub-block of the whole ${\cal M}\times {\cal M}$
scattering matrix ${\cal S}$ and can be written in the standard form, see \cite{verb85,soko89,fyod97}:
\begin{equation}\label{scattmat}
S(E,\gamma_0)=\mathbf{1}-2i W^{\dagger}\,\frac{1}{E\mathbf{1}-{\cal H}_{eff}}\,W=\frac{\mathbf{1}-iK_A}{\mathbf{1}+iK_A},
\end{equation}
where $K_A=W^{\dagger}\frac{1}{E\mathbf{1}-H+i\Gamma_{A}}W$ and we defined an effective non-Hermitian Hamiltonian as
\begin{equation}\label{scattmatt1}
 {\cal H}_{eff}=H-i\left(\Gamma+\Gamma_{A}\right), \, \quad \, \Gamma=WW^{\dagger}\ge 0, \quad \Gamma_A={\cal A}{\cal A}^{\dagger}\ge 0
\end{equation}
By using the identity $\det{(\mathbf{1}-PQ)}=\det{(\mathbf{1}-QP)}$ one can further find from (\ref{scattmat}) that
\begin{equation}\label{scattmatt3}
\det{S}(E,\gamma_0)=\frac{\det{\left(E\mathbf{1}-H+i\Gamma_{A}-i\Gamma\right)}}{\det{\left(E\mathbf{1}-H+i\Gamma_{A}+i\Gamma\right)}}
\end{equation}

 The real values $E$ for which $\det{S}(E,\gamma_0)$ vanishes define the energies of incident travelling waves for
which a CPA can be realized \cite{CGCS10}. In fact, to have CPA one has to impose a condition that the incident waveform $|I_{\rm CPA}\rangle$
must be realized as a linear combination of channel modes with amplitudes given by the components of the eigenvector of the scattering matrix
$S(E^{\rm CPA},\gamma_0)$ associated with a zero eigenvalue. Indeed, in this case $S(E^{\rm CPA}, \gamma_0)|I_{\rm CPA} \rangle=0=|O\rangle$
implying that there is no outgoing wave, hence CPA. The condition of zero eigenvalue in turn implies $\det{S(E,\gamma_0)}=0$.

The equation (\ref{scattmatt3}) makes it clear that the {\it complex} zeroes of $\det S(E,\gamma_0)$ are eigenvalues of the non-Hermitian Hamiltonian
${\cal H}^{(a)}_{eff}=H+i\left(\Gamma-\Gamma_{A}\right)$. As long as absorption is non-vanishing, i.e. $\Gamma_{A}\ne 0$,  those eigenvalues are situated in both positive and negative half-planes of the complex energy $z=E+iY$. When studying their distribution in the complex plane one can exploit the statistical rotational invariance of the Hamiltonian matrix $H$ and safely replace the matrix $\Gamma=WW^{\dagger}$ by a diagonal one $\gamma \sum_{l=1}^M \left|l\right\rangle\left\langle l|\right.$, where $\left|l\right\rangle$ is assumed to be the eigenbasis of $WW^{\dagger}$.  In addition, the physical assumption of no direct overlap between the scattering channels and the position of the absorbers allows us to consider the matrix $\Gamma_{A}$ to be diagonal as well. In the simplest representative example case $L=1$  the matrix ${\cal A}$ is rank-one, and denoting the eigenvector corresponding to its only non-vanishing eigenvalue as $\left. |0\right\rangle$  we finally see that zeroes $z_n=E_n+iY_n$ of $\det S(E,\gamma_0)$ are given by the $N$ eigenvalues of the effective non-Hermitian Hamiltonian \cite{Tsampikos}
\begin{equation}\label{1}
 {\cal H}^{(a)}_{eff}=H+iWW^{\dagger}-i\gamma_0\left.|0\right\rangle\left\langle 0|\right..
\end{equation}
 The local point absorber is modeled by the last term in (\ref{1}), with $\gamma_0>0$ standing for the loss- strength  and the
vector $\left.|0\right\rangle$ (normalized as $\left\langle 0|0\right\rangle=1$ and characterizing the position of the absorber) is assumed to be orthogonal
to all channel vectors $\left.|l\right\rangle$.

In what follows we will describe the statistics of the rescaled imaginary parts of complex zeros $y_n=\beta\pi Y_n/\Delta=\pi \beta N
\rho (E) Y_n$ as a function of the relative strength of the absorber $a\equiv\gamma_0/\gamma$. Here $\Delta =  (\rho(E)
N)^{-1}$ is the mean spacing between neighboring real eigenvalues for $H$ close to the energy $E$ and $\rho(E)=\frac{1}{\pi}
\sqrt{1-E^2/4}$ is the standard  RMT density of eigenvalues of $H$. Our main object of interest is the probability density of the {\it imaginary parts} for complex  zeroes (whose real parts are close to the energy $E$)
defined as ${\cal P}_M(y)=(N\rho(E))^{-1}\left \langle \sum_n\delta\left(y -y_n\right)\delta(E-E_n)\right\rangle_H$.\\

{\bf Perturbative treatment in the weak-coupling regime.} Before presenting the non-perturbative
results, valid for arbitrary coupling and absorption strength,
we briefly consider the limit of weak coupling $\gamma, \gamma_0\ll 1$ where the form of ${\cal P}_M(y)$ can be found
via a standard perturbative analysis, cf. \cite{FyoSav12, FyoSav15}.  In this regime one expects the zeroes to be located in the vicinity of the real axis in such a way that their 'height' in the complex plane satisfy $Y_n \ll \Delta$.
First order perturbation theory results in replacing $E_n\to z_n=E_n+iY_n$ with
\begin{equation}\label{2}
Y_n=\left\langle n|WW^{\dagger}|n\right\rangle-\gamma_0 \left|\left\langle n|0\right\rangle\right|^2=\gamma
\sum_{l=1}^M\left|\left\langle n|l\right\rangle\right|^2 -\gamma_0 |\left\langle n|0\right \rangle|^2
\end{equation}
where $ \left |n\right\rangle$ stand for the eigenvectors of $H$. In the following we assume that $E_n$ is close to $E=0$, so that the
appropriately rescaled variables are $y_n=\beta N Y_n$. The probability density ${\cal P}_M(y)=\left\langle \delta\left(y
-y_n\right)\right\rangle_H$ in this case can be further evaluated by using the fact that the projections $\left\langle n|l\right\rangle=u_l+i(\beta
-1)v_l$ of the eigenvectors of $H$ in any arbitrary basis, for large $N\gg 1$, can be effectively replaced with a set of independent,
identically distributed mean-zero Gaussian-distributed variables $u_l,v_l$ with variance $1/\beta N$.

Performing the averaging over the eigenvector components, the distribution of the rescaled imaginary parts
can be after some manipulations represented as
\begin{equation}\label{6}
{\cal P}^{(\beta)}_M(y)=C_M^{(\beta)}\,\frac{ e^{-\frac{y}{2\gamma}}}{2\gamma}\int_{-\frac{y}{2\gamma_0}}^{\infty}dt
\left(at+\frac{y}{2\gamma}\right)^{\beta\frac{M}{2}-1} \frac{e^{-(a+1)t}}{t^{1-\beta/2}} \theta(t)
\end{equation}
where $C_M^{(\beta)}=\frac{1}{\Gamma(\beta/2)\Gamma(\beta M/2)}$,
the relative strength of the absorber is conveniently characterized by the ratio $a\equiv\gamma_0/\gamma$, and $\theta(x)$ stands here and henceforth for the Heaviside step function.  Eq. (\ref{6})
is a generalization of the famous Porter-Thomas distribution for $M$ open channels, see \cite{FyoSav15} and references therein, and is reduced to it in the limit $a\to 0$.

Equation (\ref{6}) allow us to evaluate the probability $\tilde{\Pi}_M(a) \equiv\int_{-\infty}^0{\cal P}(y)\,dy$ that, for given $a$ and
$M$, a zero of the $S$-matrix crosses the real axis. One finds
\begin{equation}\label{7beta2}
\tilde{ \Pi}^{(\beta=2)}_M(a)= \left(\frac{a}{1+a}\right)^M, \quad
{\tilde \Pi}^{(\beta=1)}_M(a)=1-2\Gamma\left(\frac{M+1}{2}\right)C_M^{(\beta=1)}\int_0^{\arctan{\frac{1}{\sqrt{a}}}} (\cos{\theta})^{M-1}d\theta.
\end{equation}
for  $\beta=2$ and $\beta=1$, correspondingly. In the small $a$-limit the latter becomes
\begin{equation}\label{7b}
 \tilde{\Pi}^{(\beta=1)}_M(a)= K_M a^{M/2}+\ldots, \quad K_M=\frac{\Gamma\left(\frac{M+1}{2}\right)}{\Gamma(1/2)\Gamma\left(1+\frac{M}{2}\right)}.
\end{equation}
Equation (\ref{7b}) is relevant for CPA protocols, which usually are concerned with situations where the absorber is weak enough to absorb by itself the
incident waves.

Next let us consider a spectral strip in the complex energy plane parallel to the imaginary axis, with a width $W$around $E=\mbox{Re}z=0$, so that it contains on average
$L=W/\Delta$ complex zeroes. The number of
"perfect absorbing" states for a given set of parameters is given by the number  $L_a$ of those zeroes which are below the real axis.
  This number has the binomial distribution $\small{{\cal P}^{(\beta)}(L_a)=\left(\begin{array}{c} L_a\\ L\end{array} \right)  \left(\tilde{\Pi}^{(\beta)}_M(a)\right)^{L_a}\left(1-\tilde{\Pi}^{(\beta)}_M(a)\right)^{L-L_a}}$ characterized by the mean value  $\left\langle L_a\right\rangle=L\tilde{\Pi}^{(\beta)}_M(a)$. Specifically for $\beta=1$, we can use the above to calculate the probability ${\cal P}^{(\beta=1)}(L_a=0)$ of having no "perfect absorbers" at all in the $a\ll 1$ limit:
\begin{equation}
\label{8b}
{\cal P}^{(\beta=1)}(L_a=0)\approx \left(1-K_M a^{M/2}\right)^L\approx e^{-K_M La^{M/2}}
\end{equation}
which allow us to identify the scaling of the minimal $a=\gamma_0/\gamma$ associated with at least one CPA state.  Setting $L\sim N$
and requesting that ${\cal P}(L_a=0)=O(1)$ we get that $a_{\rm min}\sim N^{-2/M}$. The large value of $a_{\rm min}$ for $M\gg1$ is a direct
consequence of the fact that, in this limit, the density of zeroes for $\gamma_0=0$ is strongly depleted. Hence one needs larger values of $\gamma_0$
in order to drive such zeroes through the real axis.\\

{\bf Non-perturbative treatment of the problem.} For a general coupling/absorption strengths beyond perturbation regime the density ${\cal P}_M(y)$ can be inferred by a judicious re-interpretation
of the RMT results for {\it resonance} statistics obtained originally in \cite{fyod96}, and further elaborated in \cite{fyod97,somm99,fyod99}. A
fully controlled {\it ab initio} derivation is relatively involved even for the simplest $\beta=2$ case, and will be published elsewhere \cite{FyoPop}.

As is well-known \cite{verb85,soko89,fyod97}, as long as $N\gg M$ the main control parameters of the theory are the non-perturbative coupling
constants $1\le g <\infty$ and $1\le g_0<\infty$ defined in terms of 'bare' values $0\le \gamma<\infty$ and $0\le \gamma_0<\infty$ as
\begin{equation}
\label{beta2nonperturb0}
g=\frac{1}{2\pi\rho(E)}\left(\gamma+{1\over \gamma}\right);\quad g_0=\frac{1}{2\pi\rho(E)}\left(\gamma_0+{1\over \gamma_0}\right)
\end{equation}
Correspondingly, for $\beta=2$ the density of complex zeroes turns out to be given  by
\begin{equation}\label{beta2nonperturb1}
{\cal P}^{(\beta=2)}_M(y)=\frac{1}{2}\frac{\theta(-y)e^{g_0y}+\theta(y)e^{-gy}\sum_{l=1}^M \frac{[(g_0+g)y]^{l-1}}{\Gamma(l)}}{(g+g_0)^M}\int_{-1}^1e^{-y\lambda}(g+\lambda)^M(g_0-\lambda)\,d\lambda
\end{equation}
One can easily calculate the mean imaginary part $\left\langle y\right\rangle$  for the zeros:
\begin{equation}\label{MS2}
\left\langle y\right\rangle=\int_{-\infty}^{\infty} {\cal P}^{(\beta=2)}_M(y) y\,dy=\frac{\beta}{2}\left[\frac{M}{2}\ln{\frac{g+1}{g-1}}-\frac{1}{2}
\ln{\frac{g_0+1}{g_0-1}}\right].
\end{equation}
where in the latter form the expression is also valid for $\beta=1$ case considered later on.
This formula provides a generalization of the well-known {\it Moldauer-Simonius} relation, see \cite{fyod96,fyod97} and refs. therein. The
logarithmic divergence of $\left\langle y\right\rangle$ for $g\to 1$ or $g_0\to 1$ reflects the characteristic power-law tails $ {\cal P}^{(\beta
=2)}_M(y)\sim y^{-2}$ in the upper/lower half plane typical for the 'perfect coupling' case.

 The following remark is due here.  Looking at the invariance of coupling constants Eq. (\ref{beta2nonperturb0}) with respect to changing $\gamma \to 1/\gamma$ and $\gamma_0\to 1/\gamma_0$ one may naively think that there is a complete equivalence of the regimes of small and large $\gamma$'s. The situation is  however more subtle.  If both parameters $\gamma$ and $\gamma_0$ take values in the interval $0\le \gamma,\gamma_0<1$
the distribution (\ref{beta2nonperturb1}) is asymptotically exact for all $N$ zeroes.   At the same time, for $\gamma>1$ exactly $M$ 'broad zeroes'   separate and move far from the real axis in the upper half-plane, with their imaginary parts being non-random and
 given by $Y^{upper}= \gamma-\gamma^{-1}=O(1)$, see \cite{fyod97}. The overwhelming majority of $N-M$ zeroes  stay however close to the real axis at a distance of order of $1/N$ and their density is still faithfully described by the same distribution (\ref{beta2nonperturb1}).
   Such a restructuring is natural to call  the {\it zeroes self-trapping phenomenon} \cite{Tsampikos}, and its analogue
for the case of resonances is well-known and was observed experimentally \cite{trapexp}. Similarly, for $\gamma_0>1$  a single 'broad' zero with imaginary part $ Y^{lower}= -\left(\gamma_0-\gamma_0^{-1}\right)$ emerges in the lower half plane.

Use of Eq. (\ref{beta2nonperturb1}) allow us to calculate the probability $\tilde{\Pi} \equiv\int_{-\infty}^0{\cal P}_M(y)\,dy$ that, for given values of
$M, g$ and $g_0$,  a zero of the $S$-matrix crosses the real axis and hence be located in the lower half-energy plane:
\begin{equation}
\label{beta2nonperturb4}
\tilde{\Pi}^{(\beta=2)}(g,g_0)=\frac{(g+1)^{M+1}-(g-1)^{M+1}}{2(M+1)(g+g_0)^M},
\end{equation}

From Eq. (\ref{beta2nonperturb4}) we immediately infer that whenever the absorber is not {\it 'perfectly tuned'}, so that $g_0 <1$, the probability
$\tilde{\Pi}^{(\beta=2)}$ decays {\it exponentially} with the number of open channels $M\gg 1$ -- in a way similar to the perturbative regime. This is a consequence of the 'zeroes depletion', or 'gap',  arising for many open scattering channels in the upper
half-plane close to real axis. A strikingly different behavior is observed in the case of ``perfectly tuned'' absorber $g_0=1$
when the probability to have zeroes below the real axis takes the form $\tilde{\Pi}^{(\beta=2)}(g,g_0=1)=\frac{g+1}{2(M+1)} \left[1- \left(\frac{g-1}{g+1}\right)^{M+1}\right]$ and decays only as $1/(M+1)$, {\it irrespective} of the coupling strength $\gamma$ of the
cavity with the scattering channels. In particular, a spectral window around $E=0$ with a large enough number of levels ($> M$) will on average contain
a zero below the real axis. Hence, a single 'perfectly tuned' absorber dramatically increases the probability of realizing a CPA state. In fact, even in the
case of slight detuning $g_0=1+\frac{\delta}{M}$ (where $\delta\sim {\cal O}(1)$ and $M\to \infty$) we get from  (\ref{beta2nonperturb4}) $\tilde{\Pi}(g,g_0)\sim\frac{g+1}{2(M+1)}e^{-\delta/(g+1)}$.

Re-interpreting the results of \cite{somm99} one may also provide exact non-perturbative expressions for the distribution of zeroes for $\beta=1$.
In this case, however, the formulas are quite cumbersome, even for the simplest $M=1$ case. In the latter case the density of zeroes in the {\it upper}
half-plane takes the form
\begin{equation}
\label{beta1nonperturb1}
{\cal P}^{(\beta=1)}_{M=1}(y>0)=\frac{1}{4\pi} \int_{-1}^1\, d\lambda (1-\lambda)^2(g+\lambda)(g_0-\lambda)e^{-2\lambda y}{\cal F}(\lambda)
\end{equation}
with $${\cal F}(\lambda)=\int_g^{\infty}\frac{dp_1e^{-yp_1}}{\sqrt{p_1^2-1}\,(\lambda+p_1)^2}\, \frac{1}{\sqrt{(p_1-g)(p_1+g_0)}}
\,\int_1^{g}\frac{dp_2e^{-yp_2}}{\sqrt{p_2^2-1}\,(\lambda+p_2)^2} \, \frac{(p_1-p_2)(p_1+p_2+2\lambda)^2}{\sqrt{(g-p_2)(p_2+g_0)}}\,.$$
The density of zeroes for the lower half plane $y<0$ is obtained by replacing  $y\to -y$ in the density Eq. (\ref{beta1nonperturb1}) and exchanging
$g \leftrightarrow g_0$ everywhere.

In the limit of a single perfect channel $g=1$ we can further proceed and evaluate the probability $\tilde {\Pi}_1(g=1,g_0) =1-\int_0^{\infty}{\cal P}(y)\,dy$
that for a fixed absorptive coupling value $1\le g_0\le \infty$ a given zero of the scattering matrix becomes negative:
\[\tilde {\Pi}_1(g=1,g_0)=1-\frac{1}{4\sqrt{2(g_0+1)}} \int_1^{\infty}\frac{dp}{\sqrt{(p+1)(p+g_0)}}\]
 \begin{equation}
\label{beta1nonperturb5}
\times \left[(g_0-1)\ln{\frac{p+1}{p-1}}+(4p+3g_0+1)\left(p\ln{\frac{p+1}{p-1}}-2\right)\right]
\end{equation}

Finally, adapting the methods of  \cite{FyoSav15} one can show that the exact non-perturbative distribution (\ref{beta1nonperturb1}) takes the form (\ref{6}) in the weak coupling limit $g\gg1, g_0\gg 1$.


{\bf Simulations.} We have tested the validity of the RMT predictions via direct diagonalization of the effective RMT Hamiltonian Eq. (\ref{1}) and by an exact
search of zeros of the scattering matrix describing an actual wave chaos system i.e. a complex network of one-dimensional waveguides coupled together via
splitters. The latter system has been established during the last years as a prototype model for wave chaos studies \cite{kott00,scars} while its experimental
realization in the microwave domain has been demonstrated by a number of groups \cite{BYBLDS16,HLBSKS12,HBPSZZ04,AGBSK14}.

The RMT model can be used to describe (in the coupled mode theory approximation \cite{CMT}) a network of coupled cavities (acoustic/microwave or optical)
or LC circuits which are randomly coupled with one another. The random coupling can be introduced by the randomness in the distances between the cavities
(or by a random capacitive coupling in the case of LC circuits). The time-reversal invariance can be violated via magneto-optical effects in the case of electromagnetic
network resonators \cite{ST91} or via gyrators in the case of LC circuits \cite{LFLVK14}. The breaking of time-reversal symmetry is much more challenging in
the acoustic domain but recent developments have demonstrated that it can be achieved by incorporating circulating fluid elements \cite{acoustics1,acoustics2}.

\begin{figure}[h]
\includegraphics[width=1\columnwidth,keepaspectratio,clip]{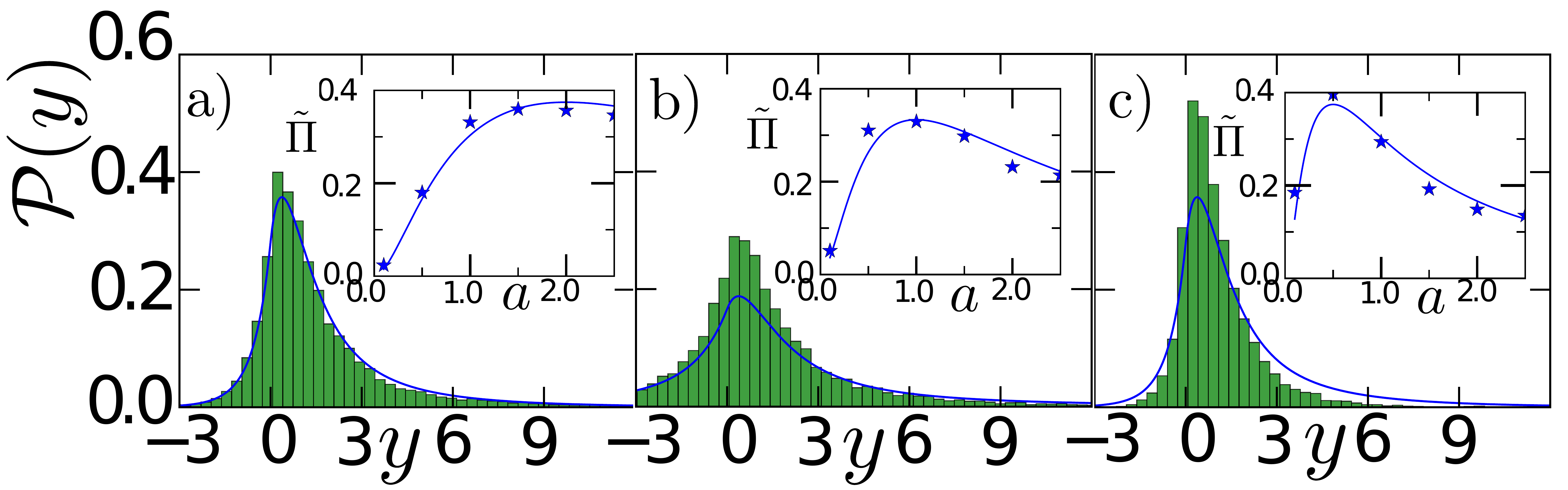}
\caption{ (color online) Distribution of $P(y)$ for a GUE RMT model with $M=2$ and $N=600$ (a) $\gamma=0.5, a=0.5$; (b) $\gamma=1, a=1$;
and (c) $\gamma=2, a=2$. In all cases the solid lines are the theoretical predictions of Eq. (\ref{beta2nonperturb1}). In the insets we report an overview
of our results for the probability $\tilde{\Pi}(\gamma,a)$ vs. $a=\gamma_0/\gamma$. The solid lines are the theoretical predictions
of Eq. (\ref{beta2nonperturb4}). The symbols are the results of our simulations.
}
\label{fig1}
\end{figure}

\begin{figure}[h]
\includegraphics[width=1\columnwidth,keepaspectratio,clip]{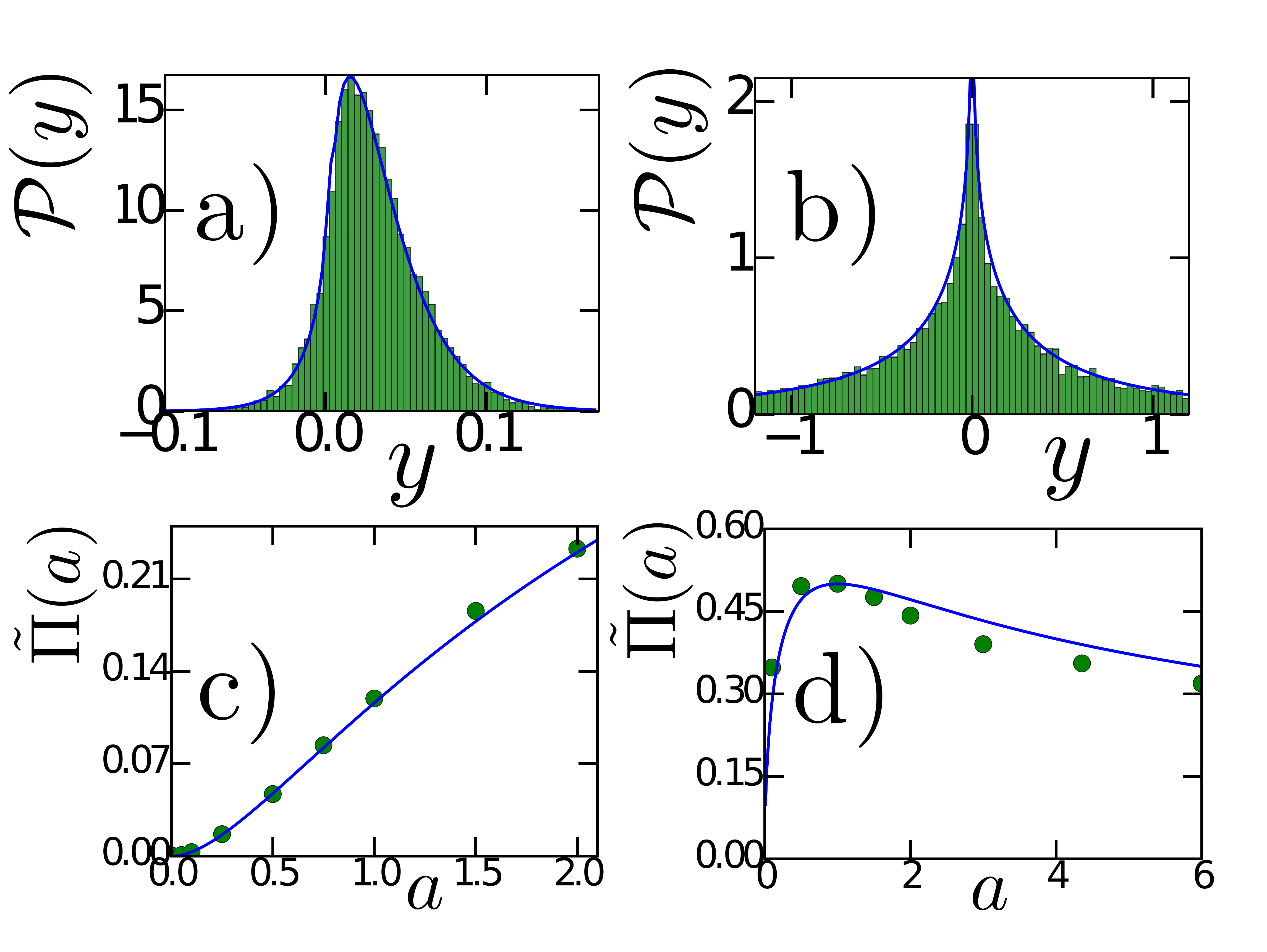}
\caption{(color online)  Distribution of zeroes $P(y)$ for (a) $\gamma=0.01,a=1$ and (b) $\gamma=1, g=1, a=1$. The solid lines represent the theoretical predictions of
Eqs. (\ref{6}) and (\ref{beta1nonperturb1}) respectively. In the lower row we report the probability of negative zeros $\tilde{\Pi}(\gamma,a)$ versus $a$ for a GOE RMT
model with (c) weak ($\gamma=0.01, M=4, N=100$) and (d) strong ($\gamma=1, g=1, M=1, N=300$) coupling. The symbols represent simulations while the lines are
the theoretical predictions Eqs. (\ref{7beta2}) for $\beta=1$ and (\ref{beta1nonperturb5}) respectively.
}
\label{fig2}
\end{figure}

We start with the presentation of the RMT simulations. We have considered a small window around $E\approx 0$ and we have collected at least $5000$
$S$-matrix zeroes for statistical processing. In Figs. \ref{fig1}a,b,c we show the distribution of zeros ${\cal P}(y)$ for a GUE ($\beta=2$) case with $M=2$ number of channels
and matrices of the size $N=600$.  The
agreement between the numerical data and the theory is excellent when the zeroes self-trapping effect is absent (see Figs. \ref{fig1}a,b), for $0<\gamma,
\gamma_0\le 1$. For $\gamma=2, a=2$ Eq. (\ref{beta2nonperturb1}) the predictions of RMT are still describing our data quite well on the scale of $1/N$,
see Fig. \ref{fig1}c, in agreement with the discussion of the resonance trapping phenomenon. In the insets of Figs. \ref{fig1}a,b,c we summarize our results for
the probability $\tilde{\Pi}^{\beta=2}(\gamma,a)$ for fixed $\gamma$-values. The numerical analysis of $\tilde{\Pi}^{\beta=1}(a)$ for the GOE case
is shown in Figs. \ref{fig2}a,b and compared with Eqs. (S\ref{6}) and (\ref{beta1nonperturb1}) for weak and strong coupling respectively. Representative probability
densities ${\cal P}(y)$ are shown in the insets of Fig. \ref{fig2}.

Next we proceed with the numerical analysis of zeroes for quantum graphs. The system consists of $n=1,\cdots, V$ vertices connected by $B$ bonds. The
number of bonds emanating from a vertex $n$ is the valency $v_n$ of the vertex and the total number of bonds can be expressed as $B={1\over 2}\sum_n v_n$.
The length of the bonds $l_{n,m}$ are taken from a uniform box distribution $l_{n,m}\in \left[l_0-w/2,l_0+w/2\right]$ where $l_0=1$ is its mean and $w=1$.
On each bond, the component $\Psi_{nm}$ of the total wavefunction is a solution of the free Helmholtz equation $\left(i{d\over dx} +A\right)^2 \Psi_{nm}(x)=
k^2\Psi_{nm}(x)$ where we have included a "magnetic field vector" $A=A_{nm}=-A_{mn}$ which breaks the time-reversal symmetry. At the vertices (where scattering
events occur), we can introduce a $\delta$-potential with a strength $\lambda_n$. Furthermore the wavefunction $\Psi_{n,m}$ at the vertices must be continuous
and must satisfy appropriate current conservation relations. We will be assuming that losses are concentrated in only one vertex which will have a potential strength
with a negative imaginary part $-i\lambda_0^{\prime\prime}$. The system is turned to a scattering set-up once leads are attached at some of the vertices. In
this case the system is described by a scattering matrix $S(k,\lambda_0^{\prime\prime})$ where $k$ is the wavenumber of an incident wave (for further details
see \cite{Tsampikos,kott00}). The zeros of the $S(k,\lambda_0^{\prime\prime})$ matrix are then evaluated numerically.

\begin{figure}[h]
\includegraphics[width=1\columnwidth,keepaspectratio,clip]{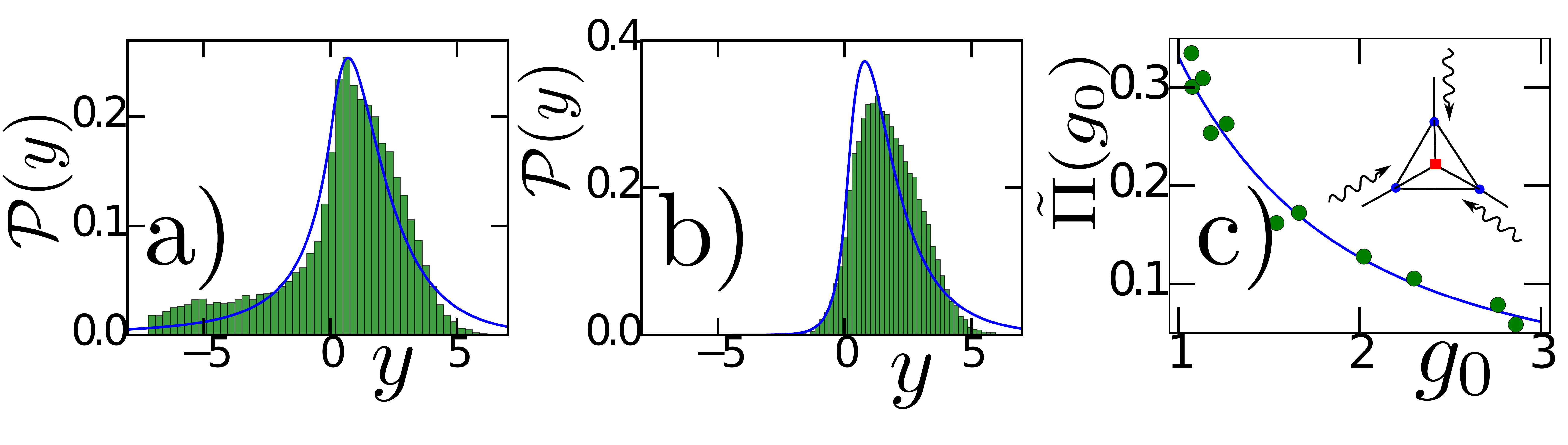}
\caption{(color online) Statistics of zeroes for a GUE ($A\neq 0$) tetrahedron graph (see inset in (c)) with total number of bonds $B=6$, valency $v_n=3$ for
all vertices and $M=3$ number of channels. The coupling strength of the graph with the lead(s) was estimated (see text) to be $g\approx 1.65$. Probability
densities ${\cal P}(y)$ for (a) $g_0\approx 1.076$ corresponding to (almost) perfectly tuned absorber and (b) $g_0\approx 2.86$. The solid lines represent
the theoretical predictions of Eqs. (\ref{beta2nonperturb1}).  (c) Probability of negative zeroes $\tilde{\Pi}(g_0)$ versus $g_0$. Symbols represent the numerical
data while the solid line represents Eq. (\ref{beta2nonperturb4}).
}
\label{fig3}
\end{figure}

Our numerical data for ${\cal P}(y)$ in two typical cases for $M=3$ of a GUE tetrahedron graph is shown in the inset of Fig. \ref{fig3}a,b. The value of the
scattering parameter $g$ (associated with the real part of $\lambda$) has been extracted in two different ways from our data for graphs. The first approach
invokes
Eq. (\ref{MS2}) in the absence of any loss. In this case the last term is zero and from the numerical evaluation of $\langle y\rangle$ we extract the value of
$g$ describing the coupling strength between the leads and the graph. The second approach utilizes the relation $\left|\langle S\rangle\right|^2=
{g-1\over g+1}$ which by using the expression (\ref{scattmat}) can be easily shown to be insensitive to any degree of loss
incorporated in the network. Both approaches gave us the same value of $g\approx 1.65$. Then the absorption coupling $g_0$ (describing the loss-strength
$\lambda_0^{\prime\prime}$) has been extracted using
Eq. (\ref{MS2}) -- this time in the presence of absorption. In Fig. \ref{fig3}a we report the case with $g_0\approx 1.07$ corresponding to (almost)
perfectly tuned absorber while in Fig. \ref{fig3}b we report the data associated with $g_0\approx 2.86$. From the figures it is obvious that in the former case
we have a proliferation of negative zeroes as predicted by Eqs. (\ref{beta2nonperturb4}). The effect is amplified even more
in the case of chaotic graphs where the numerical data for $y<0$ are consistently above the RMT prediction Eq. (\ref{beta2nonperturb1}), see Fig. \ref{fig3}a.
We speculate that the origin of this discrepancy is associated with the absence of a statistical gap in the resonance width distribution due to the presence of
scar states, see Ref. \cite{kott00,scars}. A summary of our results for ${\tilde \Pi}^{(\beta=2)}$ and various $g_0$ values are shown in Fig. \ref{fig3}c.

{\bf Conclusions.} We have investigated the statistics of complex zeroes of a scattering matrix describing a chaotic cavity with a single point-like lossy defect.
Our approach assumes that the cavity is coupled with $M$ leads and it is described by a RMT model. Using non-perturbative calculations we were able to evaluate
the density of imaginary parts of $S-$matrix zeroes. This allowed us to calculate the probability of such a zero to move to the negative part of the complex
energy plane as the loss-strength at the defect increases. We have tested our predictions both against direct numerical calculations within the RMT model and
with a dynamical system described by a complex networks of coupled waveguides. Our results are "universal" and provide new insights into the problem of chaotic
coherent perfect absorbers. Specifically they are directly addressing the question of the optimal amount of loss needed to realize a chaotic coherent perfect absorber.
It will be interesting to extend these studies beyond universality and identify dynamical effects (like scars etc) that can further enhance the efficiency of such chaotic
cavities to act as CPAs. Finally, we note that perhaps the easiest way to experimentally test our predictions for $S-$matrix complex zeroes is to use that in the framework of RMT such zeroes are statistically identical to zeroes of the diagonal $(M-1)
\times (M-1)$ sub-block of the  $M\times M$ unitary scattering matrix without absorbers. Thus we expect our results for $M=1$ should describe complex zeroes
of the $S_{11}$ element of the two-antenna device, which may be accessible by the same harmonic inversion technique used earlier in \cite{kuhl08} for extracting the resonance poles.


{\bf  Acknowledgements} (Y.V.F) was supported by  EPSRC grant  EP/N009436/1. (S.S \& T.K) acknowledge partial support from an AFOSR MURI grant FA9550-14-1-0037 and an NSF EFMA-1641109.


\vspace{\baselineskip}


\end{document}